\documentclass[fleqn,usenatbib]{mnras}

\usepackage{newtxtext,newtxmath}

\usepackage[T1]{fontenc}

\DeclareRobustCommand{\VAN}[3]{#2}
\let\VANthebibliography\thebibliography
\def\thebibliography{\DeclareRobustCommand{\VAN}[3]{##3}\VANthebibliography}

\usepackage{graphicx}	
\usepackage{amsmath}	
\usepackage{float}
\usepackage[dvipsnames]{xcolor}

\def\msun{M_\odot}
\def\ross{\rm}
\def\anatole{\rm}

\title[Galactic model effects on LISA DNS detection]{The effects of Galactic model uncertainties on LISA observations of double neutron stars}

\author[A. Storck and R.~P.~Church]{
Anatole Storck,$^{1,2}$\thanks{E-mail: storckanatole@gmail.com}
Ross P. Church$^{1,2}$
\\
$^{1}$Department of Astronomy and Theoretical Physics, Lund Observatory, Box 43, SE-221 00, Lund, Sweden\\
$^{2}$Lund Observatory, Division of Astrophysics, Department of Physics, Box 43, SE-221 00, Lund, Sweden\\
}

\date{Accepted XXX. Received YYY; in original form ZZZ}

\pubyear{2022}

\begin{document}
\label{firstpage}
\pagerange{\pageref{firstpage}--\pageref{lastpage}}
\maketitle

\begin{abstract}
Observations of  binaries containing pairs of neutron stars using the upcoming space-based gravitational wave observatory, LISA, have the potential to improve our understanding of neutron star physics and binary evolution. In this work we assess the effect of changing the model of the Milky Way's kinematics and star formation history on predictions of the population of double neutron stars that will be detected and resolved by LISA. We conclude that the spatial distribution of these binaries is insensitive to the choice of galactic models, compared to the stochastic variation induced by the small sample size. In particular, the time-consuming computation of the binaries' Galactic orbits is not necessary. The distributions of eccentricity and gravitational-wave frequency are, however, affected by the choice of star-formation history. Binaries with eccentricities $e>0.1$, which can be measured by LISA observations, are mostly younger than $100\,{\rm Myr}$. We caution that comparisons between different predictions for LISA observations need to use consistent star formation histories, and that the Galactic star formation history should be taken into account in the analysis of the observations themselves. The lack of strong dependency on Galactic models means that LISA detection of double neutron star binaries may provide a relatively clean probe of massive binary star evolution.
\end{abstract}


\section{Introduction}


The Laser Interferometer Space Antennae (LISA) is an upcoming space-based gravitational wave (GW) detector. LISA is set to launch sometime in the next decade and will observe a diversity of mHz-frequency gravitational sources that includes Galactic compact binaries in the mHz GW regime \citep{Amaro-Seoane2022}.  The most numerous such systems will be binaries containing two white dwarfs, but binaries containing neutron stars will also be observed and offer the possibility to constrain binary evolution, neutron-star formation and the equation of state of neutron-star material.  This work focuses on double neutron stars (DNSs).


We currently know of approximately a dozen DNS candidates, as summarised in \citet{Tauris2017}. DNSs are of particular astrophysical interest because short gamma-ray bursts {\anatole can} originate from the coalescence of DNSs: this model was corroborated by the observation of GRB 170817A and its associated kilanova shortly following the first DNS merger observed in gravitational waves, GW 170817 \citep{Goldstein2017, Abbott2017, Smartt2017, Metzger2019}. Studying DNS systems is challenging, as at least one of the neutron stars needs to be a pulsar to permit detection. DNSs are particularly bright sources of GWs, and are expected to be detected by LISA millions of years before their coalescence. LISA observations of DNSs will provide insight into the structure and post-formation behaviour of neutron stars (see \citet{Ozel2016} for a review), formation channels of DNS systems \citep{Andrews2020}, detection of binary pulsars \citep{Kyutoku2019}, and short-duration gamma-ray bursts.

In this work, we explore how various models for the Milky Way affect the distribution of LISA observable properties of DNSs. Theoretical predictions for the population of DNSs visible to LISA import uncertainties from both binary evolution and modelling of the stellar populations and dynamics of the Milky Way. The largest uncertainties likely lie in the binary evolution – for example treatment of common-envelope evolution and the winds of low-metallicity stars \citep{Vigna-Gomez2018} – and this will allow LISA observations to constrain the astrophysics of DNS formation. However, in order to quantify the impact of different binary evolution prescriptions it is necessary to understand the effect of the model adopted for the Galactic environment. Past works on LISA-resolvable DNSs have used a range of analytic models for the Galaxy (see e.g. \citet{Lau2020} and \citet{Andrews2020}) and local-group galaxies \citep{Seto2019}. \citet{Lamberts2019} used cosmological simulations from the FIRE project \citep{Hopkins2014} to predict the LISA white dwarf population.{\anatole \citet{YuJeffery2015} studied the effect of varying the galactic star formation history and binary evolution physics on the number and properties of the expected population.} None of these works, however, systematically study the effect of varying both their Galactic models and star-formation rates. Understanding the uncertainty in the LISA DNS detection rate caused by uncertainties in galactic models is valuable for understanding LISA's utility to study DNS formation and evolution.

We take a single prescription for the formation of DNSs and investigate the impact on the expected LISA-observed population of changing the Galactic potential, stellar distribution and star-formation history. In Section~\ref{sec:Fiducial} we describe our {\it Fiducial} model for the Milky Way, and the resulting LISA-visible binary population is presented in Section~\ref{sec:FidResults}. In Section~\ref{sec:AltModels}, alternate models are proposed for the Milky Way disc, Galactic potential, and star-formation rate. The effects of these alternate models on the observable population of DNSs are presented in Section~\ref{sec:Results}, and discussed in Section~\ref{sec:Discussion}.

\section{The Fiducial simulation}\label{sec:Fiducial}

In this section, we present our {\it Fiducial} model for DNS simulations in the Milky Way. We build up a population using the orbital parameters and space velocities of DNSs at the time of DNS formation from \citet{Ross2011}, and normalise the amount of DNSs to the Milky Way (§\ref{subsec:popSynth}). Birth times are assigned to the DNSs and their binary orbits\footnote{By {\it binary orbit} we denote the orbit of the two neutron stars around their common centre of mass; we refer to the orbit of the DNS binary around the Milky Way as the {\it Galactic orbit}} are integrated to the present day. DNSs are included in the population if {\anatole their initial eccentricity $e_i$ and present day second-harmonic GW frequency $f_{\rm gw, circ} \equiv 2/P_{\rm orb}$, where $P_{\rm orb}$ is the binary's orbital period,\footnote{For a circular binary all gravitational waves are emitted with frequency $f_{\rm gw, circ}$} meet the following conditions}

\begin{figure}
	\includegraphics[width=\columnwidth]{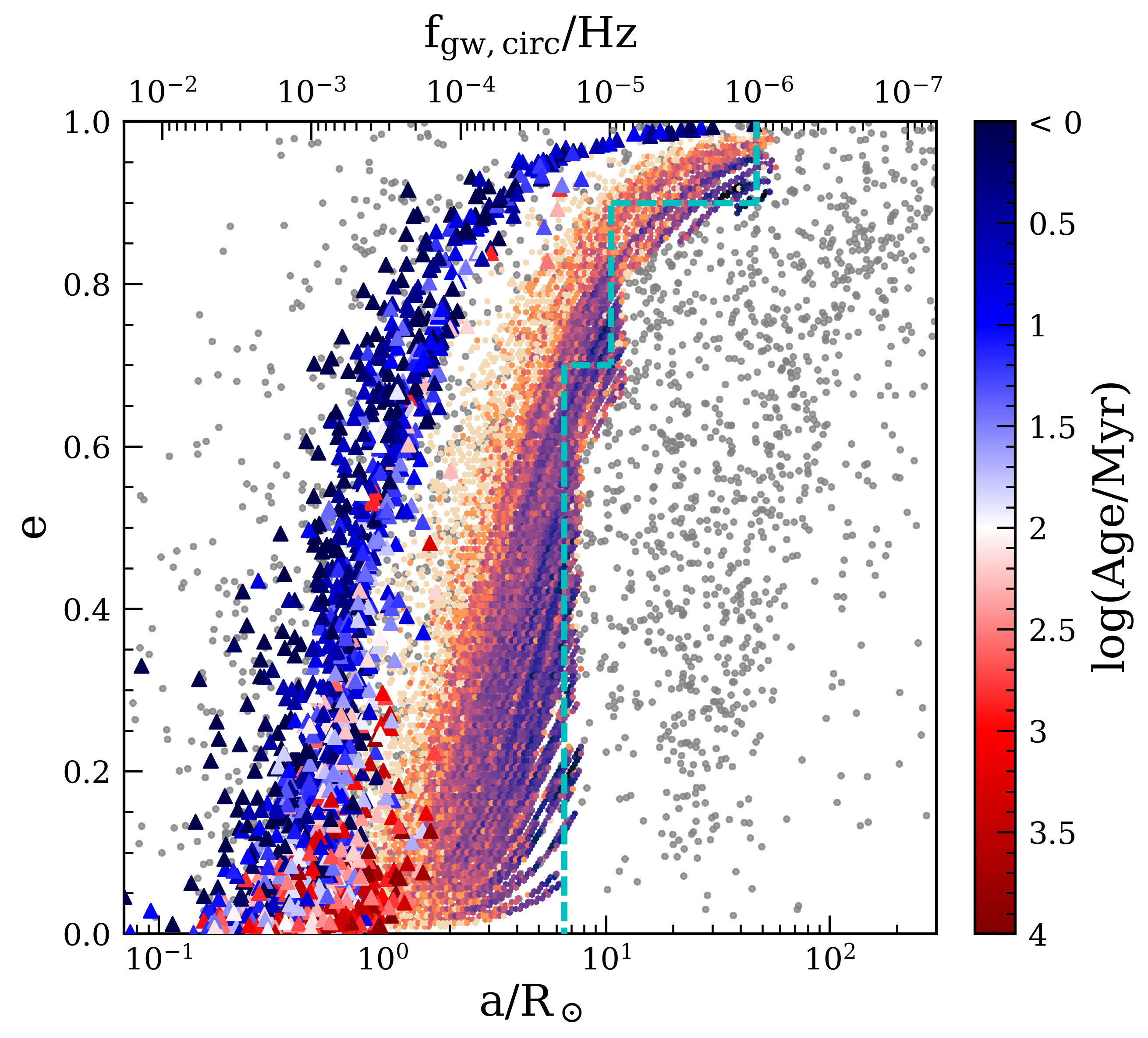}
    \caption{\ross{
    Eccentricity $e$ as a function of semi-major axis $a$ for DNS binaries in our {\it Fiducial} population.  The corresponding GW frequency $f_{\rm gw, circ} \equiv 2/P_{\rm orb}$ is on the top axis, assuming a circular DNS with two $1.4\:\text{M}_\odot$ neutron stars. Triangles show the present-day orbits of all resolvable binaries in fifty realisations of the {\it Fiducial} model, coloured according to their ages.  Grey dots show the orbits at DNS formation of all binaries from the \citet{Ross2011} input population that never appear in the {\it Fiducial} population, either because they are too short-lived (small $a$) or too wide to become resolvable (large $a$).  The shaded colourmap shows the present-day distribution of DNSs, both resolvable and not, in the {\it Fiducial} population.  The dashed cyan lines represent cutoffs we make to our input population, where any binary to the right of the lines is discarded. Some binaries appear to the right of the cutoff in the figure as they are more massive than $2.8\,\msun$ and the cut is made in $f_{\rm gw, circ}$ rather than $a$.  Tracks appear due to the resampling of the same system for different randomly assigned birth times. 
    }}
    \label{fig:AvsE}
\end{figure}

{\anatole
$$
\begin{cases}
        e_i\ge 0 & \text{for } 1>f_{\rm gw, circ}>2\times10^{-5}\,\text{Hz}\\
        e_i>0.7 & \text{for } 2\times10^{-5}>f_{\rm gw, circ}>10^{-5}\,\text{Hz}\\
        e_i>0.9 & \text{for } 10^{-5}>f_{\rm gw, circ}>10^{-6}\,\text{Hz}\\
\end{cases}
$$

since empirically binaries that lie outside these cutoffs are never resolvable. We show the cutoff criteria in Figure~\ref{fig:AvsE}.} The remaining DNSs are assigned an initial position in the Milky Way disc and their Galactic orbits are integrated to the present time (§\ref{subsec:MWmodel}). The signal-to-noise ratio is calculated for all binaries to determine the LISA-resolvable DNS population (§\ref{subsec:LISA}).

\subsection{Binary population synthesis and orbital evolution}\label{subsec:popSynth}
\label{sect:popsyn}
{\anatole
The population of DNSs is taken from \citet{Ross2011}, who synthesised a population of DNSs using the rapid binary population synthesis code BSE \citep{Hurley2002}, modified according to the core mass--remnant mass relationship of \citet{Belczynski+08}.  They evolved a population of binaries from the zero-age main sequence at solar metallicity, with both masses drawn independently from the \citet{KTG93} initial mass function above $3\,\msun$ and the orbital semi-major axis $a$ taken to be flat in $\log a$ over the range in which DNSs were produced. They drew the natal kicks of both neutron stars independently from the distribution of \citep{Arzoumanian2002}, which is on the stronger side of literature kick distributions; they found that this choice reproduced well the orbital properties of observed Galactic DNSs.  \citet{Ross2011} evolved $4\times10^7$ galactic binaries which yielded $2857$ bound DNSs; we take these as our input population. As our ``neutron-star kick'' we adopt the total effect of the natal kicks and mass-loss kicks on the Galactic velocity of the centre of mass of the DNS.}

We normalise the population to the Milky Way by matching the inferred present-day DNS merger rate of $\mathcal{R}_{\rm MW}=42\,{\rm Myr^{-1}}$ in the whole Milky Way \citep{Belczynski2018, Pol2019}. This merger rate is the peak probability value of observational data from DNS merging within a Hubble time, with a confidence of $90\%$ between $28$ and $72$ DNSs per Myr.
To calculate the merger rate per Myr of a given simulated population, we assign birth times to each DNS and calculate how many DNSs merge during the last Myr.  The merger rate varies across different realisations of the same population, so we take an average merger rate over 100 realisations.


The {\it Fiducial} model assumes a constant star-formation rate (SFR) for the Milky Way, with a maximum age of $10$ Gyr. To obtain the correct present-day Galactic merger rate we scale the \citet{Ross2011} population by a factor of {\anatole 256} by randomly re-sampling.



We integrate the orbital parameters of the DNS population by following the orbital evolution due to quadrupole GW emission. The eccentricities and semi-major axes are evolved using the following equations from \citet{Peters1964}:
\begin{align}
&\begin{aligned}
    \label{eqn:daPeters1964}
    \dot{a} = -\frac{64}{5}\frac{G^3m_1m_2(m_1+m_2)}{a^3c^5\left(1-e^2\right)^{7/2}}\left(1+\frac{73}{24}e^2+\frac{37}{96}e^4\right),
\end{aligned}\\
&\begin{aligned}
    \label{eqn:dePeters1964}
    \dot{e} = -\frac{305}{15}e\frac{G^3m_1m_2(m_1+m_2)}{a^4c^5\left(1-e^2\right)^{5/2}}\left(1+\frac{121}{304}e^2\right),
\end{aligned}
\end{align}
where $a$ is the semi-major axis, $e$ is the eccentricity, $m_1$ and $m_2$ are the masses of the two neutron stars.

\subsection{Galactic disc and potential}\label{subsec:MWmodel}

We assign positions to the DNS population by using the exponential density model of the Milky Way's stellar disc \citep{Gilmore1983, McMillan2016}. DNSs are assumed to form in the disc's mid-plane due to the short evolutionary time-scale of their progenitors. The stellar density follows
\begin{equation}
    \rho_d(R, z) = \frac{\Sigma_0}{2z_d}\exp{\left(-\frac{|z|}{z_d}-\frac{R}{R_d}\right)},
\end{equation}
with a central surface density {\anatole $\Sigma_0 = 886.7\pm116.2\,M_\odot\:\text{pc}^{-2}$, scale height $z_d=300\,$pc, and scale length $R_d=2.60\pm0.52\,$kpc.} We assume that all of the DNSs form in the thin disc.

The stellar surface density for the plane of the disc is obtained by integrating the stellar density for all $z$.
\begin{equation}
    \label{eq:stellarDensity}
    \Sigma(R) = \int_{-\infty}^{\infty} \rho_d(R, z)dz = \Sigma_0\exp{\left(-\frac{R}{R_d}\right)}.
\end{equation}
We sample the distribution in Equation~\ref{eq:stellarDensity} to build up the initial positions of the DNS population.


Using the integrator developed by \citet{McMillan2016}, we integrate the positions in the Milky Way of the DNS population from their births to present day. We use McMillan's \textit{PJM$\_$17} potential which is based on an axisymmetric density model. We assume that each DNS progenitor is in a circular Galactic orbit prior to DNS formation. DNSs receive natal kicks at their births, so the velocity of a DNS is the sum of its circular velocity and {\anatole the total centre-of-mass velocity of the DNS at formation (see Section~\ref{sect:popsyn})} which we take from \citet{Ross2011} and distribute isotropically in direction.

\subsection{Resolvability to LISA}\label{subsec:LISA}

The LISA sensitivity curve is expressed as the effective noise power spectral density $S_n$ from \citet{Robson2019}:
\begin{align}
&\begin{aligned}
    \label{eqn:LisaCurve}
    S_n(f) = \frac{10}{3L^2}\left[P_{\text{OMS}}+2\left(1+\cos^2{\left(\frac{f}{f_*}\right)}\right)\frac{P_{\text{acc}}}{(2\pi f)^4}\right]\left(1+\frac{3}{5}\left(\frac{f}{f_*}\right)\right),
\end{aligned}\\
&\begin{aligned}
    \label{eqn:POMS}
    P_{\text{OMS}} = (1.5\times^{-11}\:\text{m})^2\left(1+\left(\frac{2\:\text{mHz}}{f}\right)^4\right)\:\text{Hz}^{-1},
\end{aligned}\\
&\begin{aligned}
    \label{eqn:PACC}
    P_{\text{acc}} = (3\times^{-15}\:\text{m}\:\text{s}^{-2})^2\left(1+\left(\frac{0.4\:\text{mHz}}{f}\right)^2\right)\left(1+\left(\frac{f}{8\:\text{mHz}}\right)^4\right)\:\text{Hz}^{-1}
\end{aligned}
\end{align}
where $L=2.5\times10^9\:$m is the length of the LISA arms, $f_*=19.09\times10^{-3}\:$Hz is the transfer frequency, $P_{\text{OMS}}$ is the single-link optical metrology noise, and $P_{\text{acc}}$ is the single proof mass acceleration noise. Both $P_{\text{OMS}}$ and $P_{\text{acc}}$ are expressed in power spectral density.

The number of DNSs per frequency bin increases with decreasing $f_{\rm gw, circ}$ -- this is true for all Galactic binaries, particularly double white dwarfs which become so highly concentrated {\anatole below} 1 mHz that LISA is unable to distinguish between individual sources \citep[{\anatole e.g.}][]{Nelemans2001, Korol2017, Lamberts2019}. This effect manifests as a galactic confusion noise $S_c$ contribution to the LISA sensitivity curve \citep{Babak2021}.
\begin{equation}
    \label{eqn:BackgroundCurve}
    S_c(f) = Af^{-7/3}e^{-\left(\frac{f}{f_1}\right)^\alpha}\frac{1}{2}\left[1+\tanh\left(-\frac{f-f_k}{f_2}\right)\right],
\end{equation}
where the parameter values are given in Table~\ref{tab:LisaCurve_values}. The complete LISA curve is then the sum of $S_n$ and $S_c$. We compare the number of resolvable DNSs for a realisation of the {\it Fiducial} model calculated with the galactic background of \citet{Babak2021}, henceforth the {\it Babak background} to the galactic background from \citet{Robson2019}, the {\it Robson background}:
\begin{equation}
    S_c(f) = Af^{-7/3}e^{-f^\alpha+\beta f\sin(\kappa f)}\left[1+\tanh\left(\gamma(f_k-f)\right)\right],
\end{equation}
where the parameter values are given in Table~\ref{tab:LisaCurve_values}. The LISA sensitivity curves resulting from the two backgrounds are shown in Figure \ref{fig:PSD_Fiducial}.

\begin{table}
    \centering
    \caption{Values of the confusion noise parameters after a $4$ yr mission \citep{Robson2019, Babak2021}.}
    \label{tab:LisaCurve_values}
    \begin{tabular}{lllllll}
        \hline
        \hline
        \text{Background} & $A$ & $\alpha$ & $f_k$ & $\gamma/f_2$ & $f_1$ & $\beta$ \\
         & $\left[10^{-44}\right]$ &  & $\left[10^{-3}\right]$ & $\left[10^{-3}\right]$ & $\left[10^{-3}\right]$ & \\ \hline
        \text{Babak} & $1.14$ & 1.8 & 2.33 & 0.31 & 1.41 & \\
        \text{Robson} & $0.9$ & 0.138 & 1.13 & 1.68 & & -221  \\ \hline
    \end{tabular}
\end{table}

Following \citet{Lau2020}, we define a binary to be resolvable to LISA if it has a SNR $\bar{\rho}$ greater than 8. The sky-averaged, inclination averaged, and polarisation-averaged SNR $\rho$ is calculated for each DNS. {\anatole We use the python package {\sc legwork} \citep{LEGWORK_apjs} to facilitate SNR calculations of binaries with non-negligible eccentricities. We also use {\sc legwork} to determine the GW harmonic at which the signal has the highest SNR, which we denote by $f_{\rm gw}$. The amplitude spectral density $\sqrt{S}$ is calculated from the SNR of the DNS and PSD of LISA at the dominant frequency $f_N$

$$\sqrt{S} = \bar{\rho} \sqrt{S_n(f_N) + S_c(f_N)}.$$

}

\subsection{Fiducial Results}\label{sec:FidResults}

\begin{figure}
	\includegraphics[width=\columnwidth]{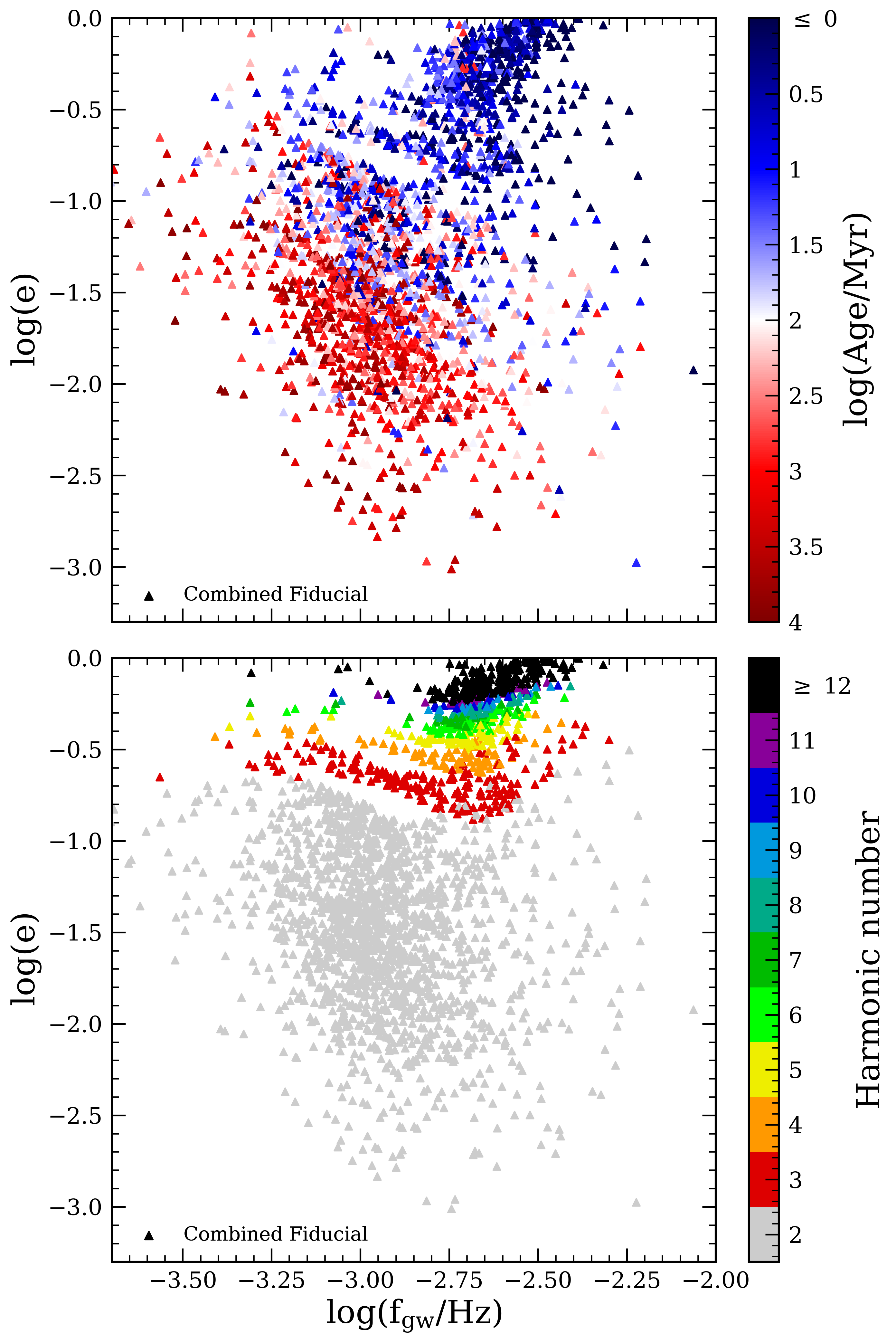}
    \caption{{\anatole
    GW frequency of resolvable DNSs from the {\it Fiducial} model accross all realisations as a function of present day eccentricity. \textbf{(top)} The DNSs are coloured according to their ages. Two distinct populations emerge: young binaries with higher eccentricities than old binaries. \textbf{(bottom)} The DNSs are coloured according to the GW harmonic which corresponds to the highest SNR.}}
    \label{fig:RB_fvse}
\end{figure}

We computed fifty realisations of the {\it Fiducial} model. 
{\ross The resolvable binaries from all these realisations are plotted as triangles in Figure~\ref{fig:AvsE}, coloured by their ages.  Young DNSs at high eccentricity can contribute even at relatively large semi-major axes since the higher harmonics of their GW emission fall in the frequency region where LISA is most sensitive.  Older resolvable binaries have had time to circularise and evolve from wide to closer orbits.  This dichotomy can clearly be seen in the upper panel of Figure~\ref{fig:RB_fvse}, where the young, high-eccentricity binaries are systematically at larger values of $f_{\rm gw}$ despite being wider than their low-eccentricity, old counterparts.  The lower panel of Figure~\ref{fig:RB_fvse} shows the frequency harmonic which contributes most strongly to the binary's detectability.  This illustrates the importance of including eccentricity when calculating gravitational-wave observables for non-circular binaries.  The background colourmap shows the population of binaries that are not resolvable from a single run of the {\it Fiducial} model.  This population increases in number towards larger orbital semi-major axis and hence we cut out the wider binaries that empirically never become resolvable to reduce the computational cost of evaluating the model; the cuts are shown as dashed cyan lines in the Figure.
}

\begin{figure}
	\includegraphics[width=\columnwidth]{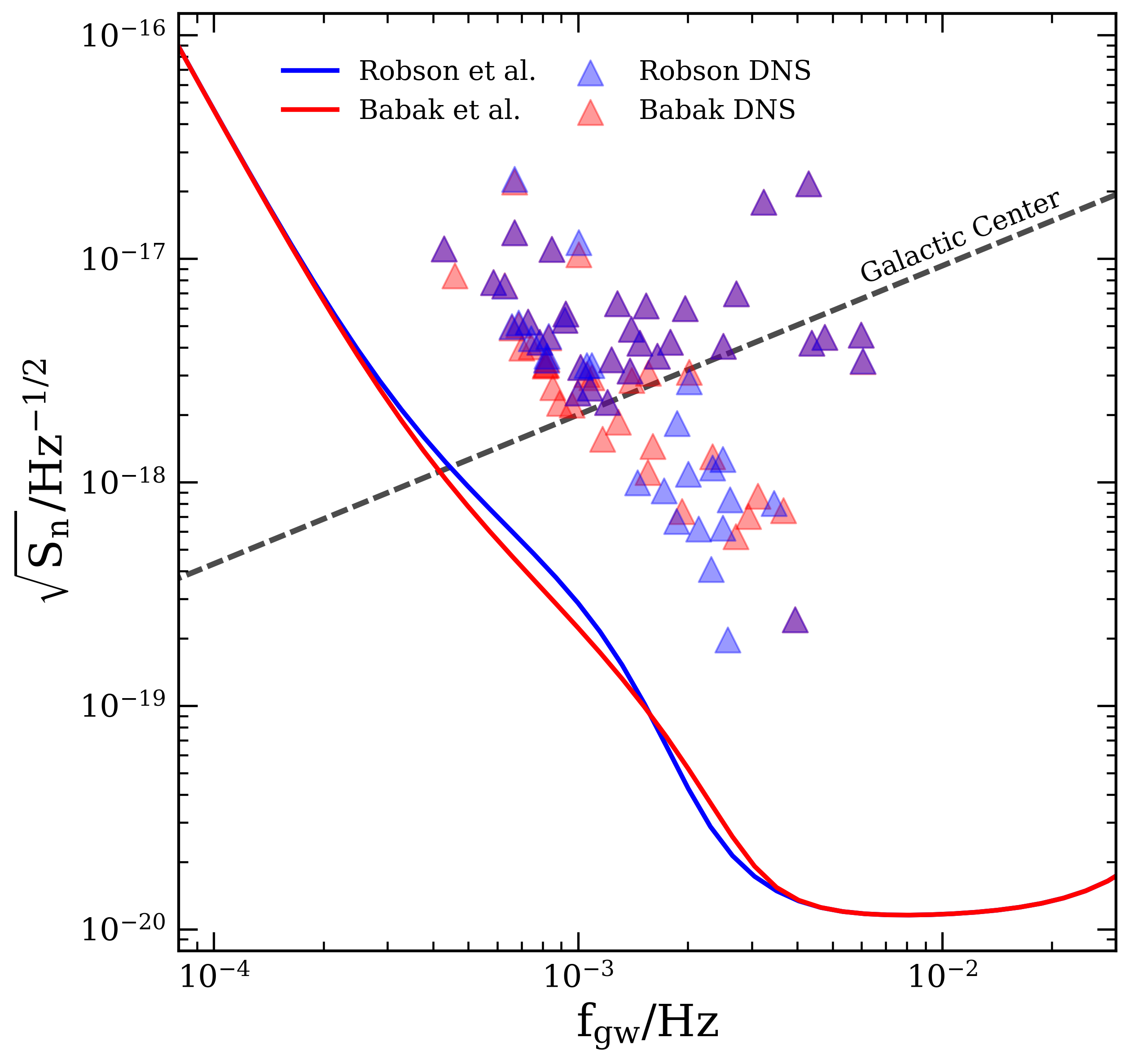}
    \caption{
    {\anatole Amplitude spectral density $\sqrt{S}$ as a function of gravitational-wave frequency $f_\text{gw}$. The blue line is the power-spectral density of noise in LISA including the Robson background confusion model; the red line is the same but for the Babak background. The blue triangles are DNSs for one realisation of the {\it Fiducial} model that are resolved by LISA (i.e. with SNR$>$8) using the Robson sensitivity curve, while the red triangles are the resolved DNSs using the Babak sensitivity curve. The DNSs are plotted according to the GW harmonic with the largest SNR. The gray dashed line is the strain amplitude of a DNS located at the distance of the Galactic Centre, 8.3\,kpc from LISA \citep{McMillan2016}. The binary is taken to be circular, with both neutron stars having masses of $1.4\,\msun$.}}
    \label{fig:PSD_Fiducial}
\end{figure}

In Figure~\ref{fig:PSD_Fiducial} we plot the {\anatole amplitude spectral density $\sqrt{S}$} as a function of $f_\text{gw}$ for a typical realisation of our {\it Fiducial} model. The number of resolvable binaries is sensitive to the SNR cut for resolvability, which in turn sensitive to the galactic background. The {\rm Robson background} peaks at lower frequencies than that of the {\rm Babak background}, towards the protrusion of DNSs going above the LISA curve. This in turn means that the {\rm Robson background} removes more of the DNSs as the curve is higher.  The number of resolvable binaries for this particular run was {\anatole 61} using the {\rm Babak background} background, and {\anatole 55} using the {\rm Robson background}. These numbers are similar to \citet{Lau2020} but distributed differently in $f_{\rm gw}$. {\anatole In particular, they are able to produce resolvable binaries in the $0.1-0.3\,$mHz range with amplitude spectral densities on the order of $10^{-16}\,\text{Hz}^{-1/2}$.}

\begin{figure*}
	\includegraphics[width=\textwidth]{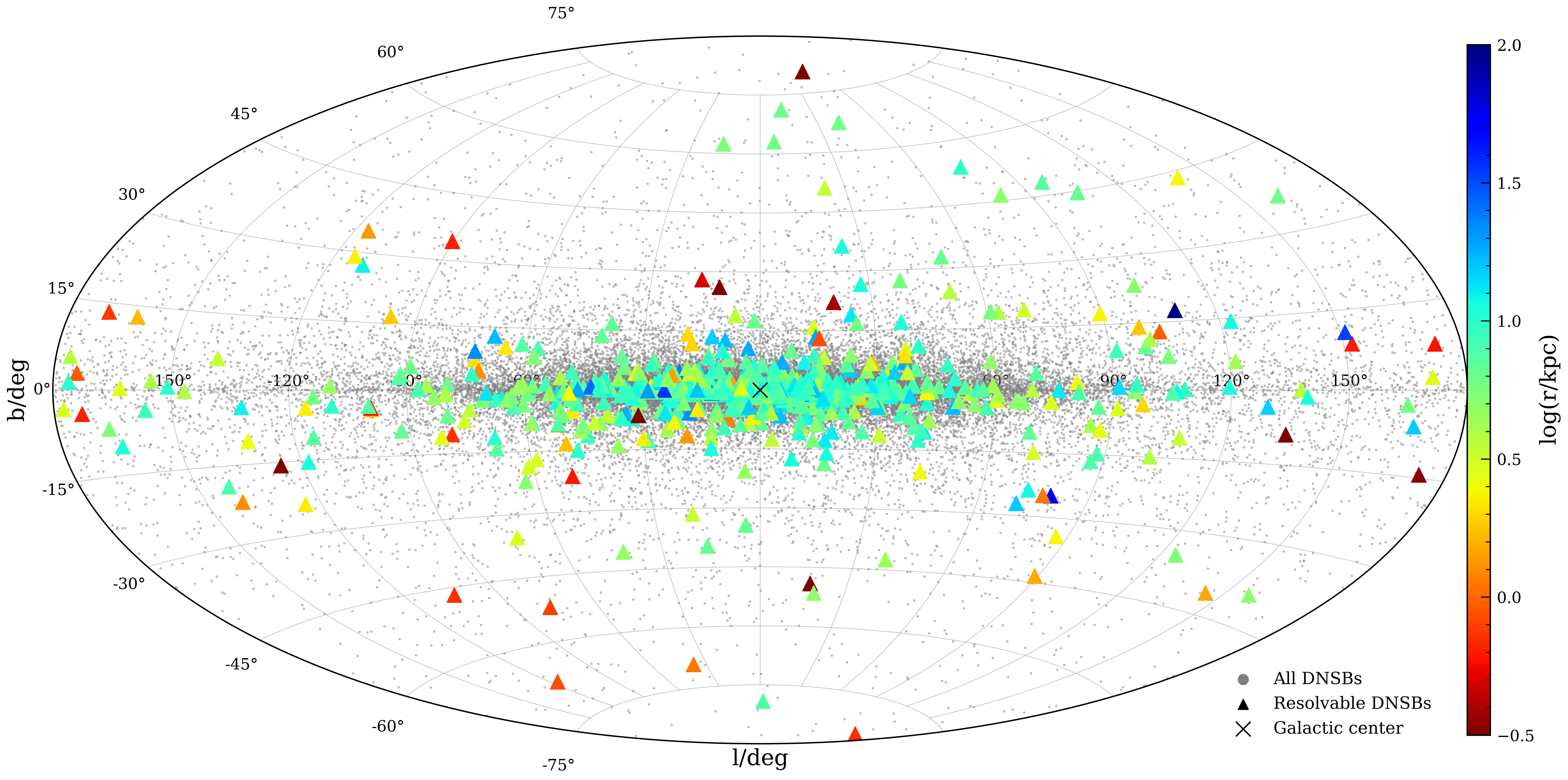}
    \caption{Sky positions in Galactic coordinates of resolvable DNSs in 15 random realisations of the {\it Fiducial} model. The DNSs are coloured according to their distance from LISA. The population of unresolvable binaries is plotted in gray. Most of the resolvable binaries are located near the Galactic Centre. Close and distant binaries can have higher latitudes: the most distant binaries are those formed with large kick velocities that overcome the galactic potential and travel far above the Galactic plane, whilst close binaries are isotropically distributed because of the vertical motion induced by the neutron-star natal kicks.}
    \label{fig:Elliptical_Fiducial}
\end{figure*}

The elliptical projection of the accumulated resolvable DNSs across fifteen realisations of the {\it Fiducial} model, as seen from the Sun, is plotted in Figure~\ref{fig:Elliptical_Fiducial}. There is a concentration of DNSs at the bulge because the region hosts the bulk of the DNSs and the potential is deep enough to retain DNSs despite their natal kicks. The local population of resolvable DNSs can be seen at all Galactic latitudes and longitudes.

\section{Alternate Models}\label{sec:AltModels}

{\ross
A summary of the set of galactic models that we consider can be found in Table~\ref{tab:modelSum}; the details of the models are presented in the following sections.
}

\subsection{Stellar density in the disc}

\subsubsection{Miyamoto distribution}\label{subsubsec:MiyaDens}

Instead of using a standalone stellar density for the disc, we can also obtain a density profile from a model potential by making use of Poisson's equation $\nabla^2\Phi_d = 4\pi G \rho_d$, henceforth the {\it MiyaDens} model. Following \citet{Lau2020}, the resulting stellar density from the \citet{Miyamoto1975} potential is
\begin{equation}
    \rho_d(R,z) = \frac{b^2_d M_d}{4\pi} \frac{a_d R^2+\left(a_d+3\sqrt{z^2+b^2_d}\right)\left(a_d+\sqrt{z^2+b^2_d}\right)^2}{\left(z^2+b^2_d\right)^{3/2} \left[R^2+\left(a_d+\sqrt{z^2+b^2_d}\right)^2\right]^{5/2}},
\end{equation}
where the constants $M_d$, $a_d$, and $b_d$ are the same as used for the potential in Section~\ref{subsec:MiyamotoPot} and can be found in Table~\ref{tab:Potential_values}. In the same way as section \ref{subsec:MWmodel}, the stellar surface density is obtained by integrating the density over all $z$ (in this case, the integration is done numerically). The stellar distribution calculated from the \citet{Miyamoto1975} potential is more extended than \citet{Gilmore1983}, with about $10\%$ of stars beyond $20\,$kpc from the Galactic Centre.

\subsubsection{Spiral arms}\label{subsubsec:Spiral}

\begin{figure}
	\includegraphics[width=\columnwidth]{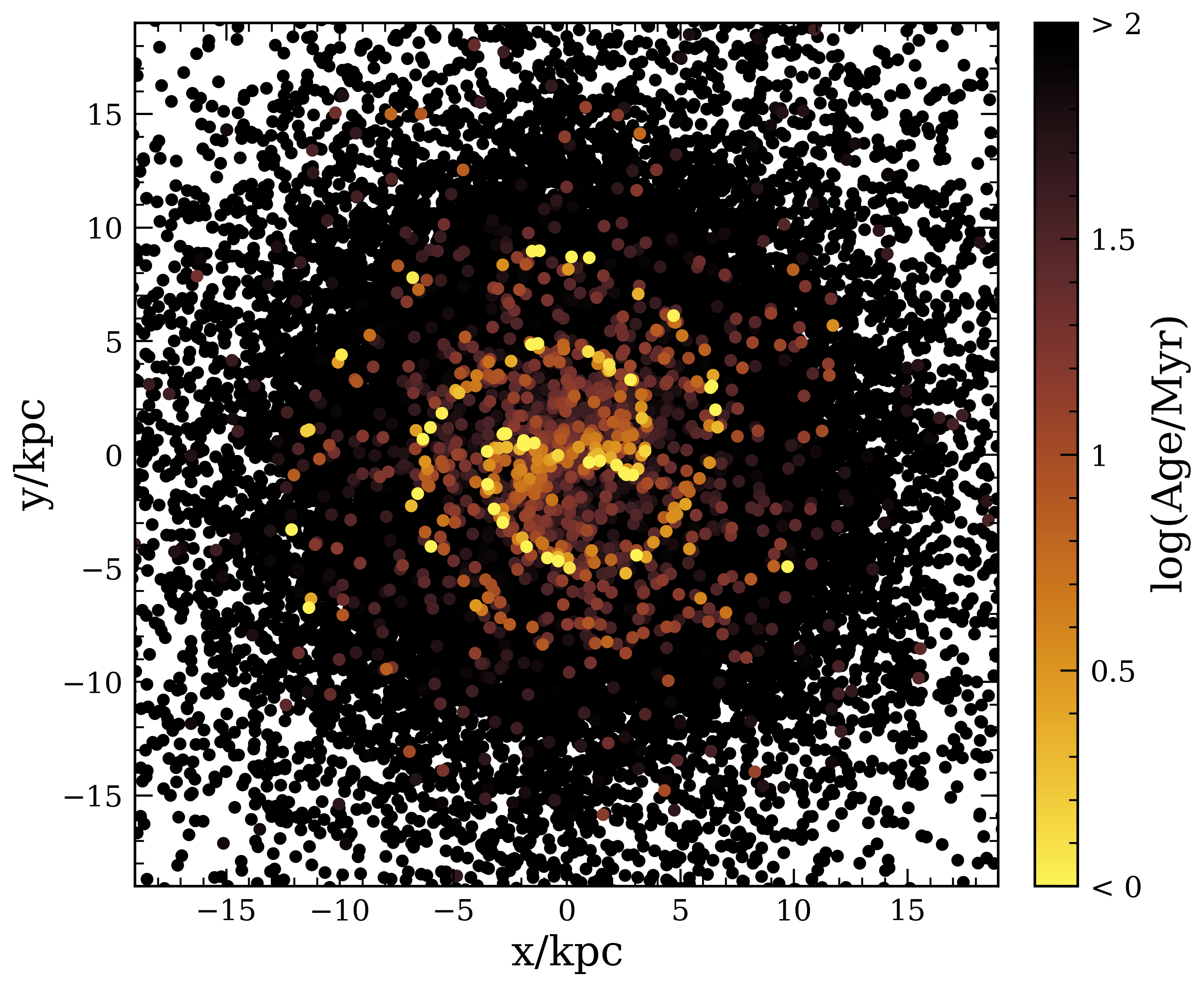}
    \caption{Top-down plot of the DNS population for one realisation of the {\it Spiral} model. The DNSs are coloured according to their birth times. A spiral structure emerges from binaries which are younger than about 20 Myr. The resolvable binary population within the spiral is a smaller subset of the full DNS population. As we see in Figure~\ref{fig:CDF_RB_d}, the spiral structure does not significantly affect the spatial distribution of resolvable binaries.}
    \label{fig:Spiral}
\end{figure}

The Milky Way's disc is not homogeneous; in particular, star formation is concentrated in the spiral arms. We consider a model where all DNSs are born in the spiral arms, henceforth the {\it Spiral} model. We use the same radial stellar density as the {\it Fiducial} model but place the binaries along the two-armed bared spiral following \citet{Pichardo2003}. The initial angle of a DNS in the {\it Fiducial} model is changed to;

\begin{equation}
    \phi_{\text{spiral}}(R) = \frac{1}{N\tan i_p}\ln{\left[1+\left(\frac{R}{R_s}\right)^N\right]} - \Omega_p t_{\text{birth}} - \frac{\pi}{6},
\end{equation}
where $N=100$ determines the prominence of the bar, $i_p=11^{\circ}$ is the winding angle, $R_s=3.3\:$kpc is the length of the bar, $\Omega_p=0.0204402\:\text{km}\:\text{s}^{-1}\:\text{kpc}^{-1}$ is the angular velocity of the spiral, and $t_{\text{birth}}$ is the birth time of a DNS. The spiral is rotated $\frac{\pi}{6}$ from the Sun-Galactic line as it is the estimated direction of the bar.

The spiral arm's angle is retroactively applied to the {\it Fiducial} model $\phi_{\text{spiral, f}} = \phi_{\text{spiral, i}} + (\phi_{\text{Fiducial, f}}-\phi_{\text{Fiducial, i}})$. This method is valid since our adopted Galactic potential is axisymmetric, hence the  angle difference between birth and present time is independent of starting angle. The effect of applying spiral arms to our population is plotted in Figure \ref{fig:Spiral}.

{\subsection{Integration models}}

\subsubsection{Miyamoto and Nagai potential}\label{subsec:MiyamotoPot}

We investigate a potential that has components from the bulge, disc, and dark matter halo \citep{Miyamoto1975, Paczynski1990}, henceforth the {\it MiyaPot} model. The potential is given in cylindrical coordinates and is axisymmetric. We choose this potential because of its analytical form and extensive use in the literature.
\begin{align}
&\begin{aligned}
    \Phi_s(R, z) = -\frac{GM_s}{\sqrt{R^2 + \left(a_s + \sqrt{z^2 + b_s^2}\right)^2}},
    \label{eqn:GalPot_s}
\end{aligned}\\
&\begin{aligned}
    \Phi_d(R, z) = -\frac{GM_d}{\sqrt{R^2 + \left(a_d + \sqrt{z^2 + b_d^2}\right)^2}},
    \label{eqn:GalPot_d}
\end{aligned}\\
&\begin{aligned}
    \Phi_h(r) = \frac{GM_c}{r_c}\left[\frac{1}{2}\ln\left(1 + \frac{r^2}{r_c^2}\right) + \frac{r_c}{r}\arctan\left(\frac{r}{r_c}\right)\right],
    \label{eqn:GalPot_h}
\end{aligned}
\end{align}
where $r=\sqrt{R^2+z^2}$. The full Miyamoto potential is the sum of the three potentials: $\Phi = \Phi_s + \Phi_d + \Phi_h$. Values for the parameters can be found in Table~\ref{tab:Potential_values}. It is shallower than the \citet{McMillan2016} potential used for the {\it Fiducial} model.


\begin{table}
    \centering
    \caption{Values used for the \citet{Miyamoto1975} potential and density.}
    \label{tab:Potential_values}
    \begin{tabular}{llll}
        \hline
        \hline
         & $M$ & $a$ & $b$   \\
         & $\left(10^{10}\:M_\odot\right)$ & $\left(10^{3}\:\text{pc}\right)$ & $\left(10^{3}\:\text{pc}\right)$  \\ \hline
        Galactic bulge ($s$) & $1.12$  & $0.0$ & $0.277$ \\
        Galactic disc ($d$) & $8.07$ & $3.7$ & $0.2$ \\
        Dark-matter halo ($h$) & $5.0$ & $6$ &  \\ \hline
    \end{tabular}
\end{table}

\subsubsection{No Integration}

Following \citet{Lau2020}, who do not integrate the positions of their DNS population but rather set their positions at the birth site of their progenitors, we include a model where the final positions of the DNS population are sampled directly from the stellar disc density given in Equation~\ref{eq:stellarDensity}. This {\it NoInt} model tests whether integrating the trajectories of DNSs has a significant effect on the population of LISA-resolvable binaries.

\subsection{Star formation rate of the Milky Way}

We investigate two alternative models for the star formation history. Both models follow an exponentially decreasing star formation rate, with one of the models including a late starburst phase.

\subsubsection{Exponential SFR profile}

We take Model A from \citet{Just2010} (henceforth the {\it J\&J} model), which uses data from the \textsc{Hipparcos} mission. The model is formulated as;
\begin{equation}
    \text{SFR}(t) = \left\langle\text{SFR}\right\rangle \frac{(t+t_0)t^3_n}{(t^2+t^2_1)^2}
\end{equation}
where $\left\langle\text{SFR}\right\rangle=3.75\:M_\odot\:\text{pc}^{-2}\:\text{Gyr}^{-1}$, $t_0=5.6\:$Gyr, $t_1=8.2\:$Gyr, and $t_n=9.9\:$Gyr. 

This model has an initially high SFR and exponentially decreases to present time, and assumes the Milky Way formed 12\,Gyr ago. To reach $\mathcal{R}_{\rm MW}=42\,{\rm Myr^{-1}}$ \citep{Belczynski2018, Pol2019}, we scale\footnote{The present-day merger rate depends largely on the current SFR. Models with a lower fraction of star formation at $t=0$ require upscaling to achieve similar present-day SFR.} the DNS population from \citet{Ross2011} by a factor of {\anatole 617.}

\subsubsection{Exponential SFR profile (with starburst)}

For our starburst model we take the SFH of \citet{Mor2019}, henceforth the {\it Mor} model, which uses data from \textit{Gaia} DR2. This model has an initially high SFR which decreases exponentially, with a starburst occurring around 2\,Gyr from the present day. \citet{Mor2019} assume that the Milky Way formed 10\,Gyr ago. The DNSs formed using the {\it Mor} model are much older than for a constant SFH, with the average resolvable DNS having a birth time of 4\,Gyr. To reach our normalisation criterion of $\mathcal{R}_{\rm MW}=42\,{\rm Myr^{-1}}$ \citep{Belczynski2018, Pol2019} we scale the DNS population from \citet{Ross2011} by a factor of {\anatole 525.}

\subsection{SFR with radial-temporal correlation}

In this model, we present a time-space coupled SFR based on an analytical inside-out model of the Milky Way from \citet{Schonrich2017}. We use this model to probe how a radial dependence on the SFR affects the characteristics of the LISA resolvable DNSs. An initial gas mass for the MW disc of $1\times10^8\:M_\odot$ is set at 12\,Gyr ago. Two gas accretion inflow components feed the galactic disc, with one component accreting $5\times 10^{10}\,M_\odot$ and the other accreting $1\times 10^{11}\,M_\odot$, supporting star formation. Both accretion components are exponentially decreasing, with exponential decay times of 1\,Gyr and 9\,Gyr respectively.

The gas disc is assumed to always follow an exponential profile, with scale length determined by
$$ R_d(t) = R_{d, 0} + (R_{d, e} - R_{d, 0})\times\left[\arctan\left(\frac{t - t_0}{t_g}\right) + \arctan\left(\frac{t_0}{t_g}\right)\right]\times N,$$
where $R_{d, 0} = 0.75\,$kpc is the initial scale length, $R_{d, e} = 3.75\,$kpc is the final scale length after 12 Gyr, and $N$ is a normalisation condition such that $R_d(t=12\,\text{Gyr}) = 3.75\,$kpc. $t_g = 2$\,Gyr is the growth time scale and $t_0 = 1$\,Gyr is the offset time.

For star formation, we follow the \citet{Kennicutt1998} law. The stellar surface density $\Sigma_\star$ changes according to the current gas density $\Sigma_g$;

$$\dot{\Sigma}_\star = 0.15 
\begin{cases}
    \Sigma^{1.4}_g & \text{if } \Sigma_g > \Sigma_\text{crit}\\
    \Sigma^{-2.6}_\text{crit} \Sigma^4_g  & \text{if } \Sigma_g \leq \Sigma_\text{crit}
\end{cases}
$$

where $\Sigma_\text{crit} = 4\,M_\odot\,\text{pc}^{-2}$ is the critical density for star formation. We assume the gas density decreases at the same rate as the stellar density increases (all of the gas is transformed into stars.)

The resulting SFH is one which resembles the {\it J\&J} model introduced in the previous section, with an initially large SFR that decreases exponentially to the present day. We make a decoupled version of the model, where we sample the radial and temporal distributions independently, to compare with our original SFH. {\anatole The coupled and decoupled models are named {\it TRadCoup} and {\it TRadDecoup} respectively. For both coupled and decoupled models, we scale our population by a factor of 670 to reach the normalisation criterion of $\mathcal{R}_{\rm MW}=42\,{\rm Myr^{-1}}$ \citep{Belczynski2018, Pol2019}.}


\begin{table*}
    \anatole{
    \centering
    \caption{Summary of each model including the stellar distribution, Galactic potential, SFR distribution. // indicates that the model is using the same component as the {\it Fiducial} model.}
    \label{tab:modelSum}
    \begin{tabular}{l|r|rrrrrrr}
        \hline
        \hline
         & Fiducial & MiyaPot & NoInt & MiyaDens & Spiral & J\&J & Mor & TRad \\
        \hline
        \vspace{0.3cm}
        \begin{tabular}[c]{@{}l@{}}Stellar\\ distribution\end{tabular} & Equation \ref{eq:stellarDensity} & // & // & \begin{tabular}[c]{@{}l@{}}Obtained\\ numerically\end{tabular} & \begin{tabular}[c]{@{}r@{}}//\\ w/ angle shift\end{tabular} & // & // & \begin{tabular}[c]{@{}l@{}}Obtained\\ numerically\end{tabular} \\
        \vspace{0.22cm}
        \begin{tabular}[c]{@{}l@{}}Galactic\\ potential\end{tabular} & PJM\_17 & Equation \ref{eqn:GalPot_d} & None & // & // & // & // & // \\
         
        \begin{tabular}[c]{@{}l@{}}SFR\\ distribution\end{tabular} & Uniform & // & // & // & // & Exponential & \begin{tabular}[c]{@{}l@{}}Exponential\\ +Starburst\end{tabular} & \begin{tabular}[c]{@{}l@{}}Obtained\\ numerically\end{tabular}\\\hline
    \end{tabular}
    }
\end{table*}

\section{Results}\label{sec:Results}

\begin{figure}
	\includegraphics[width=\columnwidth]{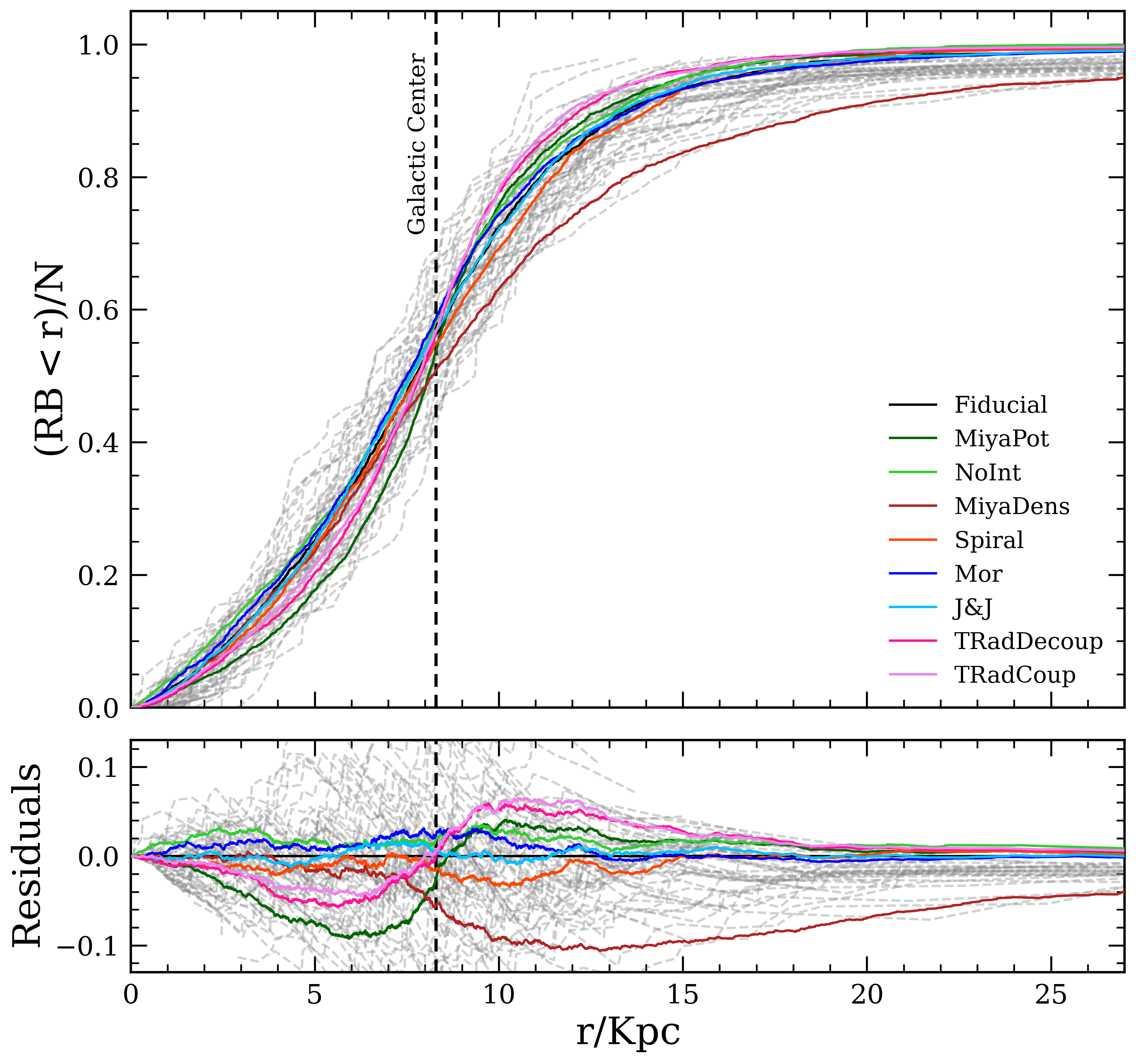}
    \caption{CDF of the resolvable DNS population as a function of distance from the Sun. CDFs for individual realisations of the {\it Fiducial} model are plotted as dashed gray lines. The residuals of alternate models to the {\it Fiducial} model are plotted below the CDF. The distance from LISA to the Galactic Centre is plotted as a vertical gray dashed line. The {\it MiyaDens} model produces binaries which are systematically further from LISA than the {\it Fiducial} model. Both {\it TRad} models produce binaries which are more clustered around the Galactic centre than the {\it Fiducial} model.}
    \label{fig:CDF_RB_d}
\end{figure}

\begin{figure}
	\includegraphics[width=\columnwidth]{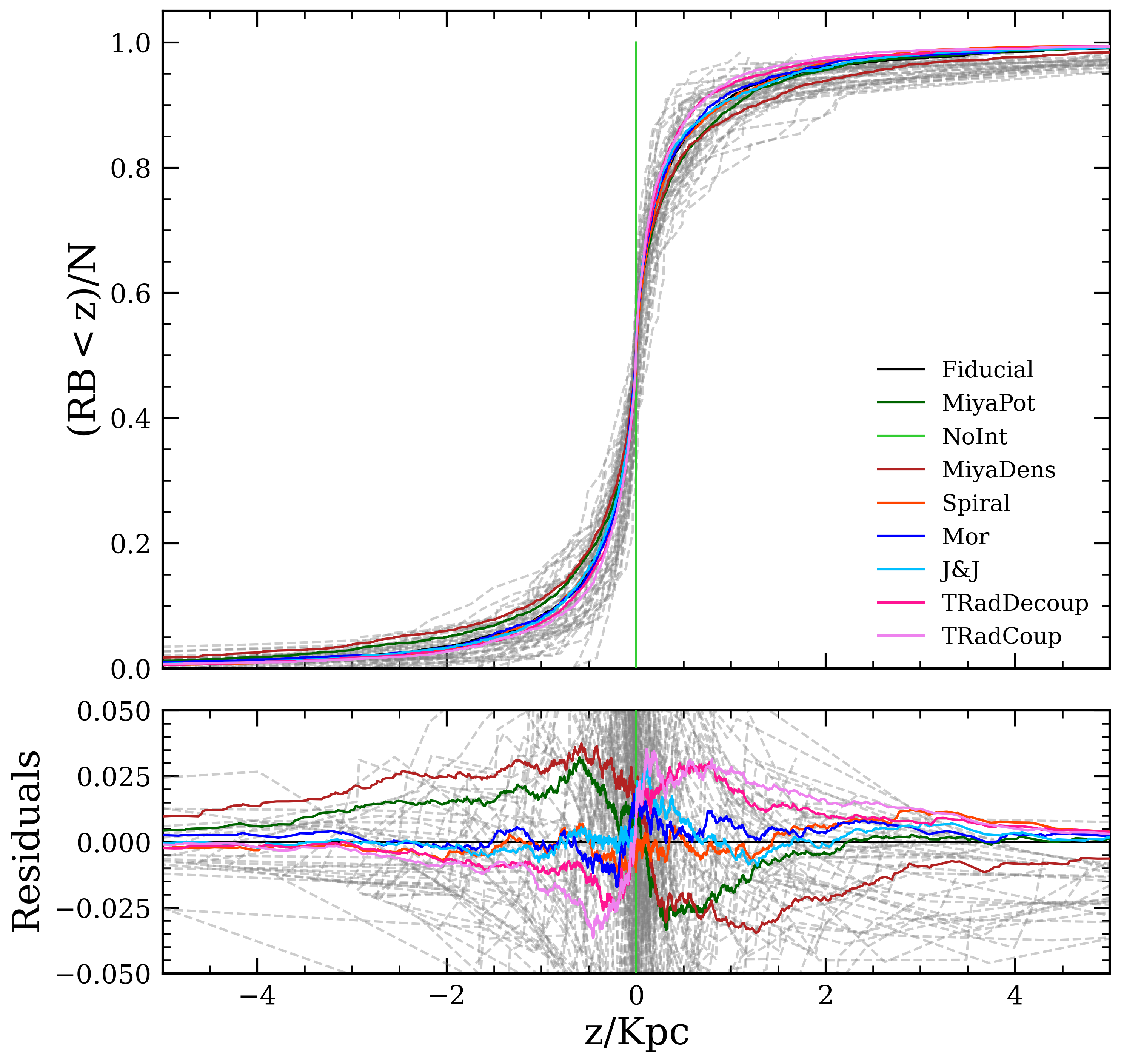}
    \caption{{\anatole CDF of the resolvable DNS population as a function of distance above the Galactic disk. CDFs for individual realisations of the {\it Fiducial} model are plotted as dashed gray lines. The residuals of alternate models to the {\it Fiducial} model are plotted below the CDF. The {\it MiyaDens} model shows the largest deviation from the {\it Fiducial} model towards a puffier disk, while the {\it TRad} models show the largest deviation towards a flat disc, apart from the {\it NoInt} model in which $z=0$ for all DNSs.)}}
    \label{fig:CDF_RB_z}
\end{figure}

\begin{figure}
	\includegraphics[width=\columnwidth]{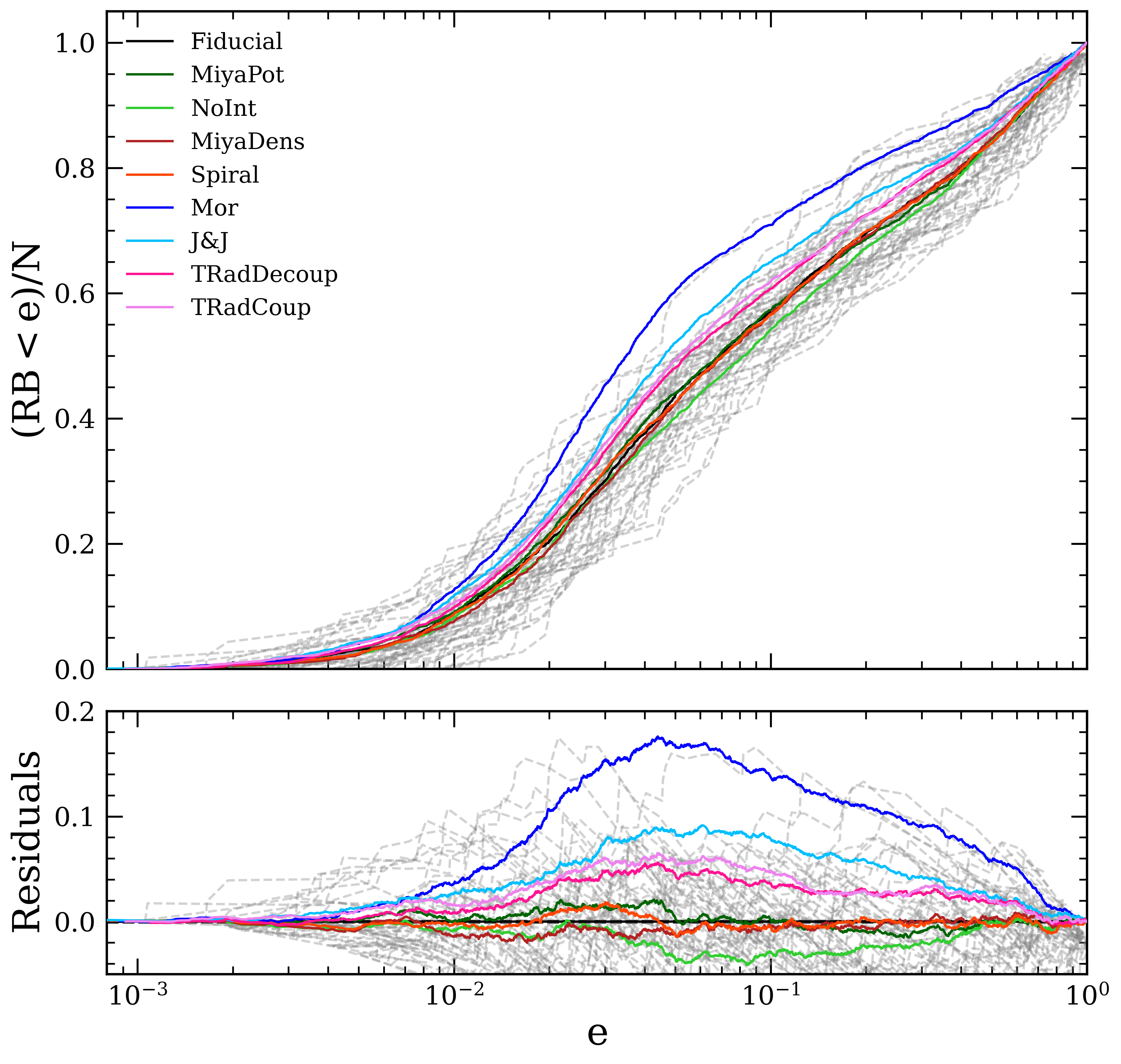}
    \caption{CDF of the resolvable DNS population as a function of their present day eccentricities. CDFs for individual realisations of the {\it Fiducial} model are plotted as dashed gray lines. The residuals of alternate models to the {\it Fiducial} model are plotted below the CDF. Alternate SFH models show a substantial increase in low-eccentricity binaries due to having systematically older populations (see Figure \ref{fig:CDF_RB_tbirth}).}
    \label{fig:CDF_RB_e}
\end{figure}

\begin{figure}
	\includegraphics[width=\columnwidth]{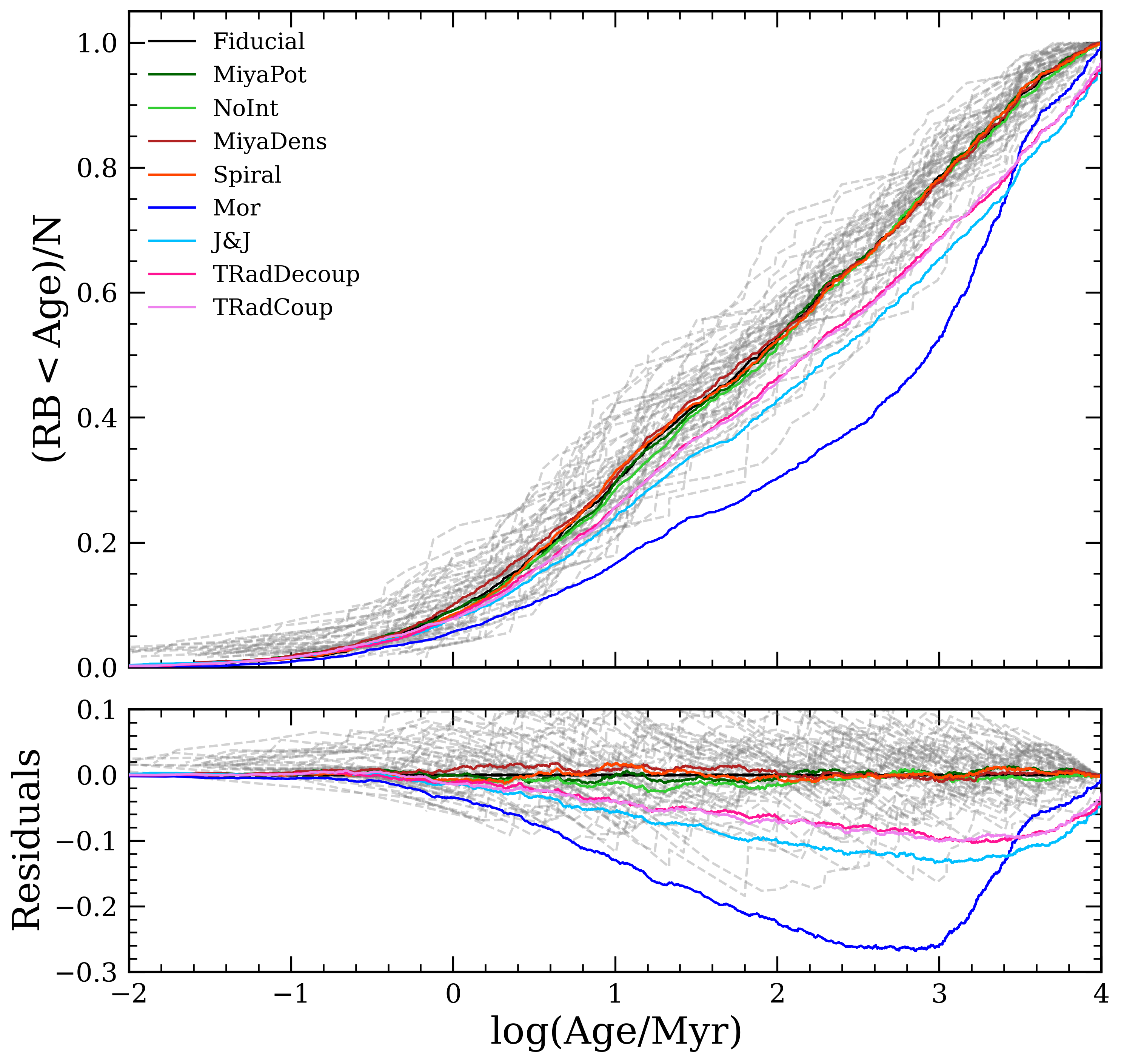}
    \caption{CDF of birth times for the resolvable DNS population. CDFs for individual realisations of the {\it Fiducial} model are plotted as dashed gray lines. The residuals of alternate models to the {\it Fiducial} model are plotted below the CDF. Alternate SFH models employ some type of exponentially decreasing SFR, producing older populations compared to the {\it Fiducial} model.}
    \label{fig:CDF_RB_tbirth}
\end{figure}

\begin{figure}
	\includegraphics[width=\columnwidth]{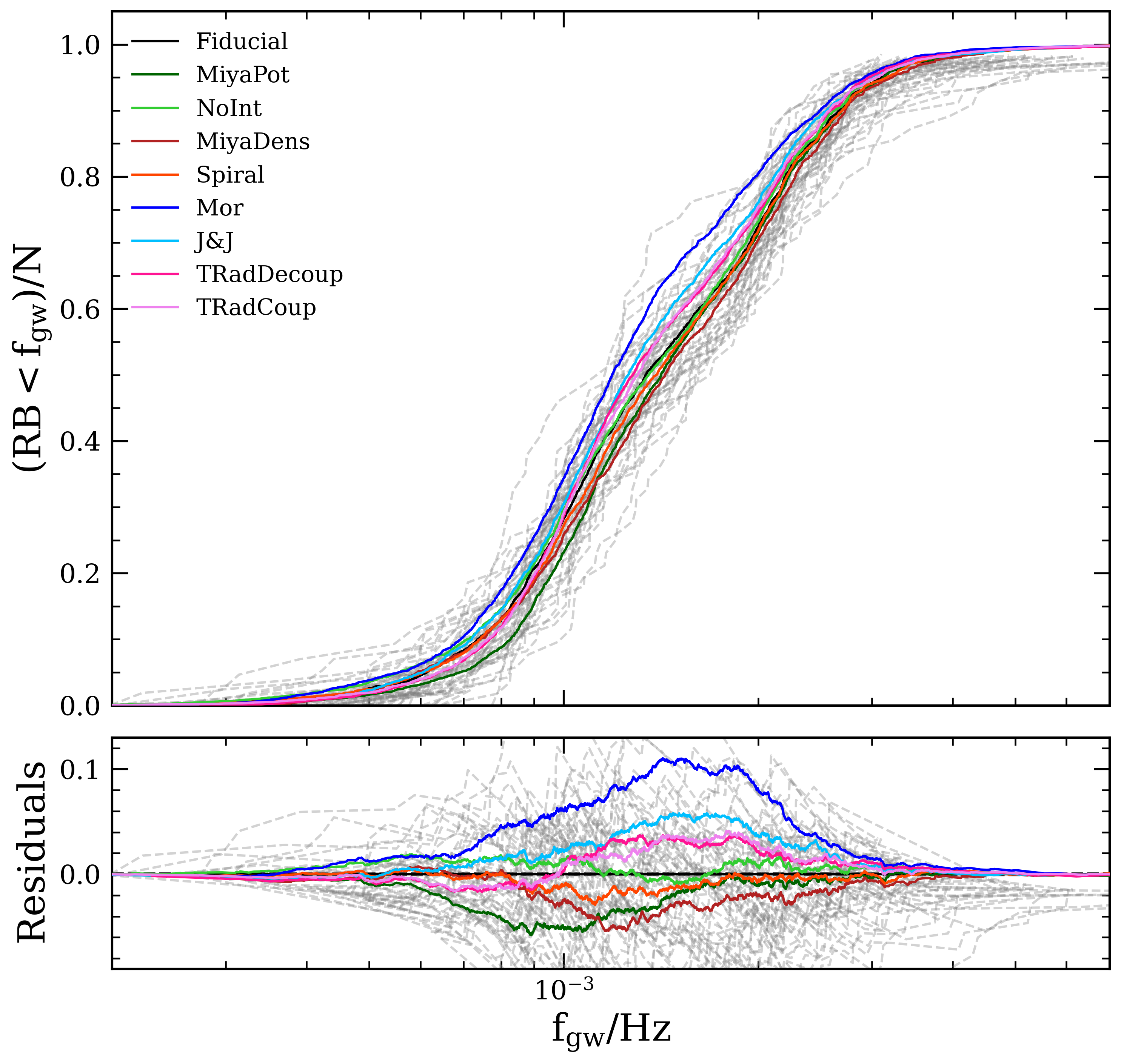}
    \caption{Present-day gravitational-wave frequency CDF of the resolvable DNS population. CDFs for individual realisations of the {\it Fiducial} model are plotted as dashed gray lines. The residuals of alternate models to the {\it Fiducial} model are plotted below the CDF.}
    \label{fig:CDF_RB_f}
\end{figure}

Fifty realisations were computed for each alternative model. The {\anatole Robson} background was used to compute the SNR. Synthetic observations are presented in the subsequent sections. We compare the distributions of distances, {\anatole both radially from LISA and distance out of the disc}, as well as 
gravitational wave frequencies 
and eccentricities 
for the resolvable DNS population across each model. 

\subsection{Number of resolvable binaries}

\begin{table*}
    \centering
    \caption{The mean $\mu$ and standard deviation $\sigma$ of the number of resolvable binaries across all runs in each model.}
    \label{tab:RBamount}
    \anatole{
    \begin{tabular}{lrrrrrrrrr}
        \hline
        \hline
         & Fiducial & MiyaPot & NoInt & MiyaDens & Spiral & J\&J & Mor & TRadCoup & TRadDecoup \\ 
        \hline
         $\mu$ & 51 & 51 & 56 & 40 & 51 & 50 & 49 & 48 & 49  \\
         $\sigma$ & 7 & 7 & 7 & 6 & 8 & 8 & 7 & 8 & 8 \\ \hline
    \end{tabular}
    }
\end{table*}

The average number of resolvable binaries across all realisations of the various models is given in Table~\ref{tab:RBamount}.  The number of resolvable binaries scales directly with the present-day DNS merger rate. Alternate determinations of the DNS merger rate show that it can be much lower than $\mathcal{R}_{\rm MW}=42\,{\rm Myr^{-1}}$ \citep{Kim2015, Chruslinska2018, Kruckow2018}; On the other hand, the empirical merger rate of the MW from the observation of two DNS mergers by LIGO is $\mathcal{R}_{\rm MW}=210\,{\rm Myr^{-1}}$ \citep{LIGO2019}. This means that the number of DNS system resolvable to LISA, while influenced by changing galactic models as shown in Table~\ref{tab:RBamount}, is dominated by the uncertainty on $\mathcal{R}_{\rm MW}$.

The average number of resolvable binaries produced varies by model, especially for models considering alternate stellar densities. The {\it MiyaDens} model has the lowest number of resolvable binaries per run. The stellar density is lower between the Galactic Centre and LISA, compared to the {\it Fiducial} model, yielding fewer binaries. Binaries near the Galactic Centre make up the majority of high-frequency systems, whereas lower-frequency binaries must be closer to LISA in order to be observed. The binaries lost from this change in stellar density are those whose frequencies are around $10^{-3}\:$Hz since their resolvability is more sensitive to changes in distance, compared to binaries with $f_{\rm gw}>6\times 10^{-3}\:$Hz.

\subsection{Spatial distribution of resolvable binaries}

A cumulative distribution function (CDF) of the distances from LISA to the resolvable binaries in our various models is shown in Figure~\ref{fig:CDF_RB_d}. Most of the alternative models are consistent with the {\it Fiducial} model given the spread of individual runs, with the exception of the {\it MiyaDens} model. The {\it MiyaDens} model extends to larger galactic radii than the \citet{Gilmore1983} distribution while the {\it TRad} models have their stars concentrated close to the galactic centre; these distributions are reflected in the distance CDFs of resolvable DNSs. Omitting the galactic integration of DNSs yields a slightly higher fraction of resolvable DNSs within $5\:$kpc, with their density around the solar neighbourhood reflecting the \citet{Gilmore1983} distribution. {\anatole In particular, the resolvable binaries beyond $22\:$kpc ($2\%$ in the {\it Fiducial} model) are removed as a consequence of not allowing the natal kicks to influence the orbits of the DNSs.} The \citet{Miyamoto1975} potential is shallower than the PJM\_17 potential from the {\it Fiducial} model, allowing DNS natal kicks to eject binaries further from the disc. As a result, a lower percentage of resolvable DNSs are within $7\:$kpc for the {\it MiyaPot} model compared to the {\it Fiducial} model. The effect, although noticeable in Figure~\ref{fig:CDF_RB_d}, is not strong enough to be discernible from a random realisation of the {\it Fiducial} model. {\anatole The CDF of distance $z$ out of the Galactic disc is shown in Figure~\ref{fig:CDF_RB_z}. All models are compatible with the {\it Fiducial} model with the exception of the NoInt model.}

\subsection{Eccentricities of resolvable binaries}

The distributions of eccentricities of resolvable binaries are plotted in Figure~\ref{fig:CDF_RB_e}. The observed distributions are model-independent with the exception of models which change the adopted Galactic star formation history ({\it Mor}, {\it J\&J}, and {\it TRad}). The different star formation histories lead to different distributions of birth times for resolvable binaries: see Figure~\ref{fig:CDF_RB_tbirth}. In particular, the exponentially decaying SFRs at the present day in the {\it Mor}, {\it J\&J}, and {\it TRad} models lead to a much lower amount of resolved DNSs with ages between 100\,Myr and 1\,Gyr. Figure~\ref{fig:RB_fvse} shows the relationship between $f_{\rm gw}$, $e$ and age for the {\it Fiducial} model. High-eccentricity binaries are almost all young – they are born visible to LISA and evolve to merge in less than 100\,Myr because their high eccentricities increase the rate at which their orbits contract. Binaries that are born wider and evolve into visibility on $\sim$Gyr timescales circularise in the process of doing so and hence have $e<0.1$ when observed. Models with a constant SFR, in which there is a significant number of young binaries, produce systems with measurable eccentricities ($e>0.1$). Finally, for low eccentricities ($e<0.1$) the evolution of $f_{\rm gw}$ is largely independent of $e$, and hence of the birth time distribution. {\anatole For the models which have a constant SFR, $\sim$45\% of DNSs have eccentricities greater than 0.1. This percentage drops down to $\sim$40\% for the {\it TRad} and {\it J\&J} models. The {\it Mor} model has the lowest percentage of high-eccentricity binaries, with less than $35$\% of resolvable DNSs having eccentricities greater than 0.1.}

\subsection{Frequency distribution of resolvable binaries}

The GW frequency distributions of most models (Figure~\ref{fig:CDF_RB_f}) are distributed the same as for the {\it Fiducial} model to within the spread of individual {\it Fiducial} runs, with the exception of the {\it Mor} model. {\anatole All SFH models exhibit a trend towards lower frequency binaries, which is a direct consequence of their lower number of high eccentricity, young DNSs. A higher eccentricity will increase the loudest harmonic up to larger frequencies (see Figure~\ref{fig:RB_fvse}). The effect is most apparent in the {\it Mor} model which has 10\% less binaries with $f_{\rm gw}>2\times10^{-3}\,$Hz.} Small deviations from the {\it Fiducial} model in alternate density and potential models occur due to the difference in spatial distributions discussed in section 5.2. Low-frequency binaries ($f_\text{gw}<10^{-3}\:$Hz) must be at distances of $d<d_\text{GC}$ to have sufficient SNR to be resolvable; see Figure \ref{fig:PSD_Fiducial}. Hence the frequency distribution depends on the relative ratio of local and distant binaries.

\section{Discussion}\label{sec:Discussion}

We investigate the observable consequences of using different galactic models when studying the LISA-resolvable population of DNSs. Our goal is to understand how sophisticated the Galactic modelling needs to be when studying the likely populations of double neutron star binaries that LISA will see.


The observable properties -- spatial distribution, gravitational wave frequency, and eccentricity -- are largely insensitive to a reasonable choice of models. Changing the initial positions of DNSs does not affect the frequency or eccentricity of resolvable binaries. An extended stellar distribution as presented in Section~\ref{subsubsec:MiyaDens} will bias the spatial distribution of binaries with $d>d_\text{GC}$ to larger galactic radii, in turn decreasing the total number of resolvable binaries; however, the overall number of resolvable binaries is a weak probe of binary evolution physics due to its normalisation to a highly uncertain empirical merger rate \citep{LIGO2019}. An alternative normalisation to the theoretically predicted merger rate derived from binary evolution and the Galactic star formation history would also come with a large uncertainty \citep{Lamberts2019}. Therefore, alternative observables such as the distribution of eccentricities, are better candidates for constraining the physics of DNS formation.

Integrating the DNSs' galactic orbits is not necessary to predict the LISA-resolved DNS population, {\anatole and the resolved population does not depend strongly on the DNS Galactic orbit calculation}. The dominant population of LISA-resolvable binaries lives near the Galactic Centre, where $|\Phi|\gg v_\text{kick}^2$ so retention of binaries in the Galaxy is likely. {\anatole The choice of galactic model does not strongly impact the distribution of distance above the Galactic plane ($z$), which suggests that this observable may provide a useful constraint on the distribution of neutron-star kicks in DNSs.} The spiral pattern from Section~\ref{subsubsec:Spiral} is only visible in binaries younger than $20\:$Myr, which is a very small fraction of observed binaries. The pattern is smeared out by differential rotation, so complete randomisation of the spiral is expected on a Galactic orbital timescale.

The Milky Way star formation history, however, matters when constructing models of the LISA revolvable DNSs. From the {\it TRad} models, we see that coupling the SFR of the MW to its stellar distribution resulted in no change to the resolvable DNS CDFs. An exponentially decreasing SFH, even including a late starburst, heavily biases the resolvable DNS population to older systems (see Figure~\ref{fig:CDF_RB_tbirth}). Gravitational wave emission then results in orbital circularization at present day (see Figures~\ref{fig:RB_fvse},~\ref{fig:CDF_RB_e}.) This means that, when predicting LISA-visible populations of DNSs, attention should be paid to the choice of star formation history. Similarly, when comparing models with different binary population synthesis approaches, care must be taken to adopt the same star formation history to avoid introducing spurious differences into the observed population. Finally, LISA observations of high-eccentricity DNSs may be a good (if noisy) probe of the total current rate of star formation Galaxy-wide, including star-forming regions that are heavily obscured by dust.

{\anatole Our results are consistent with the earlier work of \citet{YuJeffery2015}.  They studied the effect of constant and exponentially decaying star formation histories, and found that they led to very similar numbers and frequency distributions for the resolvable DNS population (see e.g. their fig. 9).  They also considered a starbust SFH which, while not realistic for the Milky Way is instructive, and found that such a SFH would lead to fewer or no resolvable DNS binaries in the present-day Milky Way, depending on the binary evolution physics.  This is broadly consistent with our Figure~9 for the {\it Mor} and {\it J\&J} models, where most of the star formation is early.  They find that differences in the frequency distribution from changing the binary evolution physics are much larger than the differences produced by changing the SFH, which implies that LISA measurements of DNS binaries have the potential to be a useful probe of binary evolution physics.}

Our conclusions come with some caveats. Despite keeping the DNS population unchanged we are unable to isolate ourselves from binary evolution models (for e.g. {\anatole SeBa} \citep{Zwart1996, Toonen2013}, BSE \citep{Hurley2002}, {\anatole COSMIC} \citep{Breivik2020}, MOBSE \citep{Mapelli2017, Giacobbo2018, Giacobbo2020}, and COMPAS \citep{Stevenson2017, Vigna-Gomez2018}), which yield various DNS birth velocity distributions. If the \citet{Ross2011} DNSs shown in Figure~\ref{fig:AvsE} have too small $v_\text{kick}$, our results may not hold. The natal kicks from \citet{Arzoumanian2002} are relatively strong so this is unlikely.

In this work we assume the ability to distinguish resolvable DNS systems from other compact binaries, specifically double white dwarfs. The chirp mass and eccentricity measurement of these systems will likely reveal the type of system if their frequency is $>1.75\times10^{-3}\:$Hz \citep{Nelemans2001, Andrews2020}. Follow up observations may be a viable option for nearby binaries since their isotropic sky positions reduce the confusion with other sources in the Galactic disc (see Figure~\ref{fig:Elliptical_Fiducial}).

\section{Conclusions}\label{sec:Conclusion}


We have built a model for the Milky Way and evolved a population of double neutron stars from formation to the present day. We have changed the stellar spatial distribution, galactic potential, and star formation histories of our models in order to investigate the observable properties of LISA-resolvable systems.

We show that varying the radial density model for the Galactic disc has a small effect on the spatial distribution of the resolvable DNSs. That in turn translates to a small shift in their frequency distribution, with no discernible change to the eccentricity CDFs. Changing the potential of the Milky Way has little effect on the spatial and frequency distributions of resolvable DNSs, and no effect on the eccentricity distributions. Lastly, we show that varying the model for the star formation rate of the Milky Way has a large effect on the birth times of the resolvable DNSs, which in turn changes the expected eccentricity {\anatole and frequency distributions}. A low present-day SFR in the MW significantly reduces the number of young detectable DNS systems. LISA will be able to distinguish binaries with $e>0.1$ from circular binaries, and we show that essentially all such binaries are the result of recent star formation.

Our results show that, when comparing different predictions for LISA DNS observations, the choice of Galactic models is relatively unimportant. However, the choice of star formation history does have a significant impact on the predicted eccentricity and frequency distributions. This should be taken into account e.g. when comparing the differences in expected LISA populations from different binary evolution prescriptions. These results suggest that LISA observations of DNS will help constrain massive stellar evolution due to the lack of effect from varying galactic models on resolvable DNS populations.

\section*{Acknowledgements}

The authors would like to thank {\anatole the anonymous referee of the original manuscript for their thoughtful and constructive comments, and} Quentin Baghi and Thomas Wagg for assistance with LISA SNR calculations, for which we used {\sc legwork} \citep{LEGWORK_joss}. The authors would also like to thank Abbas Askar, Nerea Gurrutxaga, Paul McMillan, Florent Renaud, and Álvaro Segovia for their helpful comments. This work was funded in part by the Swedish Research Council through the grant 2017-04217.

\section*{Data Availability}

The data used in this paper can be obtained on request to the corresponding author.
 



\bibliographystyle{mnras}
\bibliography{Paper2021.bib} 




\bsp	
\label{lastpage}
\end{document}